\documentclass[12pt]{article}
\usepackage{amssymb,epsf,graphicx}
\input epsf.sty
\newcommand{\be}{\begin{equation}}
\newcommand{\ee}{\end{equation}}
\newcommand{\ben}{\begin{eqnarray}\displaystyle}
\newcommand{\een}{\end{eqnarray}}
\newcommand{\sectiono}[1]{\section{#1}\setcounter{equation}{0}}

\newcommand{\ds}{\displaystyle}
\newcommand\crbig{\\\noalign{\vspace {1.5mm}}}

\begin{document}
{}~ \hfill\vbox{\hbox{hep-th/0304213}\hbox{CTP-MIT-3362} }\break
\vskip 2.1cm

\centerline{\large \bf Tachyon condensation in open-closed p-adic string
theory }
\vspace*{8.0ex}

\centerline{\large \rm Nicolas Moeller and Martin Schnabl}

\vspace*{8.0ex}

\centerline{\large \it Center for Theoretical Physics}
\centerline{\large \it Massachusetts Institute of Technology,}
\centerline{\large \it Cambridge, MA 02139, USA} \vspace*{2.0ex}
\centerline{E-mail: {\tt moeller@lns.mit.edu, schnabl@lns.mit.edu}}

\vspace*{6.0ex}

\centerline{\bf Abstract}
\bigskip
We study a simple model of p-adic closed and open strings. It sheds
some light on the dynamics of tachyon condensation for both types of
strings. We calculate the effect of static and decaying D-brane
configurations on the closed string background. For closed string
tachyons we find lumps analogous to D-branes. By studying their
fluctuation spectrum and the D-branes they admit, we argue that closed
string lumps should be interpreted as spacetimes of lower
dimensionality described by some noncritical p-adic string theory.

\vfill \eject

\baselineskip=16pt

\tableofcontents

\sectiono{Introduction}
\label{s_intro}

The study of open p-adic string theory \cite{Brekke,Brekke-Freund}
has revealed that this theory
possesses many nonperturbative features that are common to open string
field theory; yet p-adic string theory is a very simple model whose
solutions, in some cases, can be found analytically. Besides its
perturbative vacuum, it has a true vacuum
in which there is no solution to the linearized equations of
motion\footnote{Anharmonic oscillations in the true vacuum were found in
\cite{Moeller-Zwiebach}.}, a situation that is believed to happen in
open string field theory, where the open string excitations should
disappear once the D-brane condenses \cite{senconj}. P-adic string
theory also possesses solitonic lump solutions
\cite{Brekke,Ghoshal:2000dd,Joe_modes} that can be interpreted as
lower dimensional D-branes. This theory is thus a nice model for open
string tachyon condensation.

An interesting question is whether one could extend these p-adic
models to study the condensation of closed string tachyons as
well. Closed p-adic string amplitudes have been studied in
\cite{Freund-Olson-Witten}, and a closed p-adic Lagrangian was derived
in \cite{Frampton-Nishino}. This Lagrangian, however, looks exactly
like an open p-adic Lagrangian, only with a different tachyon
mass. And if one takes this model seriously, one would have to give an
interpretation for the closed string ``D-branes'' of any dimension,
that exist in this theory.  Other difficulties related to closed
strings were pointed out in \cite{Minahan-quantum}. In particular it
was shown that one-loop diagrams of open p-adic string theory do not
possess any pole that could correspond to massless closed strings.

However, in \cite{Brekke-Freund} Brekke and Freund presented a very
interesting mixed Lagrangian of both open and closed p-adic tachyons,
that was derived from the p-adic $n$-point amplitudes for open and
closed tachyons. Surprisingly, we haven't found any study of this
Lagrangian in subsequent literature. It is the object of this paper to
study this open and closed p-adic theory.

Although the pure open and pure closed p-adic string theories behave
similarly, we will see that the two fields have in fact very distinct
dynamics in the mixed theory. We will exhibit the fundamental
differences between both fields and show how this leads us to
interpret the lumps of the closed tachyon as lower dimensional
spacetimes. In particular, we will see that this picture is coherent
since we cannot have D-branes with more worldvolume dimensions than
the dimension of spacetime. We further support our claims by studying
the fluctuations of closed string lumps. All fluctuations are strictly
normalizable, which means that all excitations are confined to the
worldvolume of the lump.  Far away from the core of the lump, the
closed string tachyon is in its true vacuum, and one can check that
the linearized equations of motion do not admit any solutions.  The
fluctuation spectrum is equidistant, strongly suggesting that the full
theory around these lumps should be described by some sort of
lower-dimensional noncritical string theory. We are tempted to
speculate that similar lumps should exist in ordinary string theories,
in which case one would get localization to lower dimensions of all
physics including the gravity.

Another domain of application of open-closed p-adic string theory is
to study the effect of static or decaying D-branes on the ambient
spacetime. Again, the ambient spacetime in this theory is fully
specified by the closed string tachyon background. Of particular
interest are the time dependent solutions
\cite{Gutperle-Strominger,Sen:2002nu,Sen:2002in,Sen:2002vv}, which
have recently attracted a lot of attention and whose physics has not
yet been fully understood.  One of the most pressing problems is that
the approximations of the open cubic string field theory considered so
far \cite{Moeller-Zwiebach,Yang:2002nm,Aref'eva:2003qu,Fujita:2003ex}
failed to reproduce the results obtained by conformal field theory
arguments.  Instead of the open string tachyon relaxing in the true
vacuum, the authors found oscillations with ever growing amplitude. It
is not clear whether this is caused by the omission of closed strings,
by the crude approximation taking into account the tachyon field only,
or whether it reflects just a wrong choice of variables.

One possibility is that the string field theory calculations can be
reconciled with CFT by taking into account that at finite string
coupling the rolling open string tachyon dissipates its energy into
closed strings
\cite{Okuda:2002yd,Strominger:2002pc,Chen:2002fp,Maloney:2003ck,Lambert:2003zr,Gaiotto-Itzhaki-Rastelli}.
We study this process in our theory, but are unable to reach definite
conclusions. One of the reasons is that in the interacting theory of
closed and open strings it is impossible to separate the total
conserved energy into open and closed string parts.

\paragraph{}
This paper is structured as follows: First, in Section \ref{s_vacuum},
we study the vacuum structure of the open-closed p-adic theory. In
Section \ref{s_double_lump}, we study lumps of both the open and
closed fields, and we discuss how this leads us into our
interpretation of closed lumps as lower dimensional
spacetimes. Section \ref{s_back_reaction} is devoted to a study of the
back-reaction of open lumps on the flat space background. In
Section \ref{s_time} we extend those results to rolling
solutions. Finally, Section \ref{s_conclusions} is devoted to
discussions and conclusions.

\sectiono{Open-closed p-adic string theory and its vacuum structure}
\label{s_vacuum}

The full action for the coupled p-adic open and closed strings was
calculated in \cite{Brekke-Freund} by requiring that it reproduces
correctly all the $n$-point p-adic tachyon amplitudes involving both
open and closed tachyons. It is given by $S = \int{d^Dx {\cal L}}$,
where the Lagrangian is
\ben \label{Lagrangian}
{\cal L} &=& -\frac{1}{2g^2} \frac{p^2}{p-1} \phi p^{-\Box/2} \phi
-\frac{1}{2h^2} \frac{p^4}{p^2-1} \psi p^{-\Box/4} \psi
\nonumber\\
&& + \frac{1}{h^2} \frac{p^4}{p^4-1} \psi^{p^2+1}
+\frac{1}{g^2} \frac{p^2}{p^2-1} \psi^{p(p-1)/2} \left(\phi^{p+1} -1\right) \,,
\een
where $\phi$ and $\psi$ are respectively the open and closed tachyons,
$\Box = -\partial_t^2 + \nabla^2$ is the $D$-dimensional
d'Alembertian, and $g$ and $h$ are respectively the open and closed
string coupling constants\footnote{We are using units in which $\alpha'=\frac{1}{2}$.}.
The dimension $D$ of spacetime can be
chosen arbitrarily in this model. Note that in order to derive this
Lagrangian, $p$ was assumed in \cite{Brekke-Freund} to be a prime
integer greater than two. But the Lagrangian itself makes sense for
any integer $p$ greater than one. In fact, it even makes sense in the
limit $p \rightarrow 1$.

Let us look first at the vacuum structure of the theory. The potential is
\ben \label{potential}
V &=& \frac{1}{h^2} \left[
\frac{1}{2} \frac{p^4}{p^2-1} \, \psi^2 - \frac{p^4}{p^4-1} \,
\psi^{p^2+1}
\right. \nonumber \\
&& \left. \quad + \lambda^2 \left( {1 \over 2} \frac{p^2}{p-1} \, \phi^2
- \frac{p^2}{p^2-1} \, \psi^{p(p-1)/2} \left(\phi^{p+1} -1\right) \right)
\right] \,.
\een
We have introduced here a parameter $\lambda = h/g$, which governs the
interaction between the open and closed string sectors\footnote{One may wish to
rescale the fields as $\hat{\phi} = \phi/g$ and $\hat{\psi} = \psi/h$ so that
the kinetic terms are independent of the coupling constants. We find this quite
inconvenient since this would make the position of the perturbative vacuum
$(\phi, \psi) = (1, 1)$ coupling constant dependent as well as the profile of
some other classical solutions. Of course all physical results are independent
of the chosen normalization.}. In real string theory $h\sim g^2$ up to some
numerical proportionality factor. We assume this to be the case also in p-adic
string theory, though we shall leave the proportionality factor undetermined.
We thus find that $\lambda \sim g$ has the interpretation of the open string
coupling constant.

\begin{figure}[!ht]
\begin{center}
\input{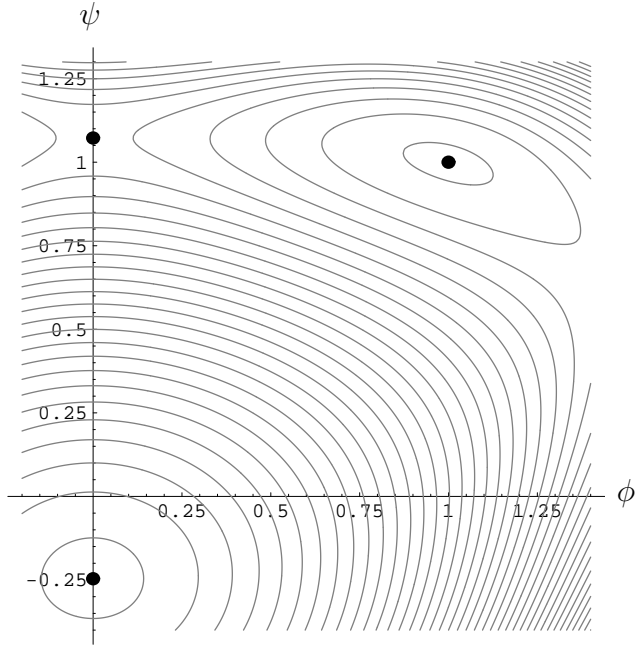}
\caption{Equipotential curves for $p=2$ and $\lambda = 1$.  The dots
mark the positions of the local minimum $(0, -0.246)$ which is the
true vacuum of the theory, the saddle point $(0, 1.07)$ which is
interpreted as a flat spacetime with no D-brane, and the local maximum
$(1, 1)$ corresponding to a D-brane filling all of spacetime.}
\label{potential_figure}
\end{center}
\end{figure}

The stationary points of the potential are given by the equations
\ben
&& \psi \left(1- \psi^{p^2-1} -
\lambda^2 \frac{p-1}{2 p} \psi^{\frac{p(p-1)}{2}-2} \left(\phi^{p+1}-1\right)
\right) = 0
\\
&&  \phi \left(1- \phi^{p-1} \psi^{\frac{p(p-1)}{2}} \right) = 0
\een
which have the following interesting solutions:
\begin{itemize}
\item
For all $p$ there is a local maximum solution $\phi=\psi=1$, this is
the perturbative vacuum for both sectors. We may interpret it as a
D-brane filling all of our flat spacetime.
\item
For $p>2$ there is always a local minimum solution with $\phi=\psi=0$,
which is supposed to be the true open and closed string vacuum. For
$p=2$ this vacuum gets shifted by the interaction to $\phi=0, \,
\psi=-\frac{\lambda^2}{4} + O(\lambda^8)$.
\item
A third interesting solution is the saddle point $\phi=0, \, \psi = 1+
\frac{\lambda^2}{2p(p+1)} +O(\lambda^4)$. It is in fact a local
minimum in the $\phi$ direction and a local maximum in the $\psi$
direction.  And it is thus interpreted as being a perturbative flat
closed string background with no D-brane.
\end{itemize}
The positions of these three extrema are shown in Figure
\ref{potential_figure} for the case $p=2$ and $\lambda = 1$.
Finally let us note that there is a couple of saddle-point solutions,
which come from infinity as the coupling $\lambda$ is turned on. We
won't be interested in these solutions here.

From this extrema analysis, we can already make an attempt to
interpret the true vacuum of $\psi$ as being a configuration in which
there is no spacetime. Indeed, we note that there is no solution
in the vicinity of $\psi=0$ and $\phi=1$, which could be a true vacuum for
the closed string and a perturbative vacuum for the open string.
This is very natural from the physical point of view,
since such a solution would represent a D-brane living outside
spacetime. We will thus adopt this interpretation, and address it
further in Section~\ref{s_double_lump}.

Finally, note that the Lagrangian (\ref{Lagrangian}) differs from the
one in \cite{Brekke-Freund} by a constant term. Indeed, the authors of
\cite{Brekke-Freund} chose to set the value of the potential to zero
at the local maximum (perturbative vacuum) $\phi = \psi = 1$. We don't
make this choice here; in fact, for all $p \neq 2$, the potential
(\ref{potential}) is zero at the local minimum $\phi = \psi = 0$.

\sectiono{Lower-dimensional spacetimes and double lumps}
\label{s_double_lump}
\newcommand{\A}{p^{-{1 \over 2} \partial_x^2}}

In this section, we will argue that lumps of the closed tachyon $\psi$
represent flat spacetime extended along the lump's worldvolume
only. Those are therefore spacetimes of lower dimensions.

\paragraph{}
Let us first write the equations of motion derived from (\ref{Lagrangian})
\be
\begin{array}{rcl} \label{eom}
\ds{p^{-\frac{1}{4} \Box} \psi} & = & \ds{\psi^{p^2} + \lambda^2 \,
\frac{p-1}{2 p} \, \psi^{\frac{p(p-1)}{2}-1} \left(\phi^{p+1}-1\right)}
\nonumber \crbig
\ds{p^{-\frac{1}{2} \Box} \phi} & = & \ds{\phi^p \, \psi^{\frac{p(p-1)}{2}}}
\,. \nonumber
\end{array}
\ee
For finite $\lambda$, these equations are very hard to solve
analytically. We will therefore start our discussion by setting $\lambda = 0$.
Recall that $\lambda$ is proportional to the string coupling.
In the limit  $\lambda = 0$ the first equation
in (\ref{eom}) determines a classical closed string background,
the second equation then tells the field $\phi$ how to propagate in that
background. The back-reaction is neglected in this limit.
For fields depending only on the $d$ spatial coordinates $x$
the equations then take the form
\ben \label{zero_eom_psi}
p^{-\frac{1}{4} \partial_x^2} \psi &=& \psi^{p^2} \\
\label{zero_eom_phi}
p^{-\frac{1}{2} \partial_x^2} \phi &=& \phi^p \, \psi^{\frac{p(p-1)}{2}} \,.
\een
Let us shortly describe the case when $\psi$ sits on its unstable
vacuum.  If we set $\psi = 1$, the equation (\ref{zero_eom_psi}) is
trivially satisfied, and equation (\ref{zero_eom_phi}) becomes exactly
the static equation of motion of open p-adic string theory, which we
know admits Gaussian lump solutions
\cite{Brekke,Ghoshal:2000dd,Joe_modes}. We thus interpret these
solutions as lower dimensional D-branes living in flat spacetime.
By flat spacetime we mean more precisely that the closed string tachyon
is in its perturbative vacuum.

We now want to consider nontrivial static configurations of $\psi$. We
readily see that (\ref{zero_eom_psi}) looks very much like the static
equation of motion of purely open p-adic string theory. And therefore
we know that it must have lump solutions. Indeed, using the identity
\be\label{identity}
e^{- a\partial^2} e^{-b x^2} =(1-4ab)^{-\frac{d}{2}}\,  e^{-\frac{b}{1-4ab}
x^2}
\ee
we find the codimension-$d$ lump solution
\be\label{double1}
\psi_0(x) = p^{d \over p^2-1} \,
e^{-{p^2-1 \over p^2 \log p} x^2}.
\ee
Far from the center of the lump $x=0$ the closed tachyon $\psi_0(x)$ rapidly approaches
its true vacuum $\psi=0$.
Assuming that $\phi_0(x)$  has finite asymptotic value and its derivatives
vanish asymptotically, it is obvious from equation (\ref{zero_eom_phi}) that
the open string tachyon itself must approach its true vacuum.
It follows then, that if we are going to find any D-brane in this background,
its dimensionality has to be smaller or equal to that of the closed tachyon lump.

In fact, such solutions can be easily found. Plugging
(\ref{double1}) in (\ref{zero_eom_phi}) and using a Gaussian ansatz
we find for example
\be\label{double2}
\phi_0(x) =
p^{d (p+2) \over 2 (p^2-1)} \, e^{-{p^2-1 \over 2 p^2 \log p} x^2}.
\ee
This corresponds to a D-brane filling out our $D-d$  dimensional spacetime.
Lower dimensional D-branes can be also easily found, but higher dimensional
ones do not exist for the reasons explained above. Since both lumps
(\ref{double1}) and (\ref{double2}) are sitting on top of each other we call
this solution a {\em double lump}.


To test our interpretation of the double lump further, let us study the
spectrum of fluctuations around it. For fluctuating closed and open string
tachyon field we write
\ben
\psi(\vec{y}, \vec{x}) &=& \psi_0(\vec{x}) + \delta\psi(\vec{y}, \vec{x}),
\nonumber\\
\phi(\vec{y}, \vec{x}) &=& \phi_0(\vec{x}) + \delta\phi(\vec{y}, \vec{x}),
\een
where $\vec{x}$ and $\vec{y}$ are the transverse and parallel coordinates to
the lump respectively. The part of the action quadratic in the fluctuations is
then
\be\label{QuadrFluct}
S_f = - {1 \over 2 h^2} {p^2 \over p-1} \int\!d^{D-d}y \, dx^d \, (\delta\psi,
\delta\phi) \left(\begin{array}{cc} K_{\psi\psi} & K_{\psi\phi} \\ K_{\phi\psi}
& K_{\phi\phi} \end{array} \right) \left(\begin{array}{c} \delta\psi \\
\delta\phi \end{array} \right).
\ee
Here
\ben\label{Kexpl}
K_{\psi\psi} &=& \frac{p^2}{p+1} p^{-\frac{1}{4}(\Box_y + \Box_x)} -
\frac{p^4}{p+1} \psi_0^{p^2-1} + O(\lambda^2),
\nonumber\\
K_{\psi\phi} = K_{\phi\psi} &=& -\lambda^2 \, \frac{p(p-1)}{2}
\psi_0^{\frac{p(p-1)}{2}-1} \phi_0^p,
\nonumber\\
K_{\phi\phi} &=& \lambda^2 p^{-\frac{1}{2}(\Box_y + \Box_x)} - \lambda^2 p
\psi_0^{\frac{p(p-1)}{2}} \phi_0^{p-1}.
\een
Following \cite{Joe_modes} we shall decompose the fluctuations as
\ben\label{HermitDecom}
\delta\psi(\vec{y},\vec{x}) &=& \sum_{n_1,\ldots,n_d =0}^\infty
u_{\vec{n}}(\vec{y}) \, H_{n_1}(\alpha x_1) \ldots H_{n_d}(\alpha x_d) \,
\psi_0(\vec{x}),
\nonumber\\
\delta\phi(\vec{y},\vec{x}) &=& \sum_{n_1,\ldots,n_d =0}^\infty
v_{\vec{n}}(\vec{y}) \, H_{n_1}(\beta x_1) \ldots H_{n_d}(\beta x_d) \,
\phi_0(\vec{x}),
\een
where $H_n(\xi)$ are the Hermite polynomials, normalized in such a way as to
obey the orthogonality condition
\be \label{ortho}
\int_{-\infty}^{\infty}{d\xi \exp(-\xi^2) H_m(\xi) H_n(\xi)} = \pi^{1 \over 2}
2^n n! \delta_{mn} \,.
\ee
Using the identity
\be
H_n(\xi) = n! \oint \frac{dz}{2\pi i} z^{-n-1} e^{-z^2 + 2z \xi},
\ee
it is quite easy to derive a generalization of the formula (\ref{identity}), which
for $d=1$ reads
\ben\label{identity2}
&& e^{- a\partial^2} H_n(\alpha x) e^{-b x^2} =
\\\nonumber
&& \qquad = (1-4ab)^{-\frac{1}{2}}\,
\left(1+\frac{4a\alpha^2}{1-4ab}\right)^{\frac{n}{2}} H_n\left(\frac{\alpha
x}{\sqrt{(1-4ab)(1-4ab+4a\alpha^2)}}\right) e^{-\frac{b}{1-4ab} x^2}.
\een
Plugging the decomposition (\ref{HermitDecom}) into the action
(\ref{QuadrFluct}) and integrating over $x$, we end up with an action for the
modes $u_{\vec{n}}(y)$ and $v_{\vec{n}}(y)$
\be\label{QuadrFl-uv}
S_f = - {1 \over 2 h^2} {p^2 \over p-1} \int\!d^{D-d}y \, \sum_{{\vec{n}},{\vec{m}}=0}^\infty
(u_{\vec{n}}, v_{\vec{n}}) \left(\begin{array}{cc}  K_{{\vec{n}}{\vec{m}}}^{uu} & K_{{\vec{n}}{\vec{m}}}^{uv} \\
K_{{\vec{n}}{\vec{m}}}^{vu}
& K_{{\vec{n}}{\vec{m}}}^{vv} \end{array} \right) \left(\begin{array}{c} u_{\vec{m}} \\
v_{\vec{m}} \end{array} \right).
\ee
To find the spectrum of fluctuations we have to look at the zero modes of the
kinetic term. Note that from (\ref{Kexpl}) we have
\be\label{detfactor}
\det \left(\begin{array}{cc}  K_{{\vec{n}}{\vec{m}}}^{uu} & K_{{\vec{n}}{\vec{m}}}^{uv} \\
K_{{\vec{n}}{\vec{m}}}^{vu} & K_{{\vec{n}}{\vec{m}}}^{vv} \end{array} \right) =
\det \left(K_{{\vec{n}}{\vec{m}}}^{uu}\right)
\det \left(K_{{\vec{n}}{\vec{m}}}^{vv}\right) + O(\lambda^2),
\ee
since to lowest order in $\lambda^2$ the determinant receives contributions only from the two
blocks on the diagonal. Therefore to lowest order the mass spectrum is given by the zeros of the
two determinants of $K_{{\vec{n}}{\vec{m}}}^{uu}$ and $K_{{\vec{n}}{\vec{m}}}^{vv}$, and is not
affected by the presence of the terms which mix the two kinds of excitations.

The two constants $\alpha$ and $\beta$, which entered our fluctuation ansatz were arbitrary.
For a convenient choice
\ben
\alpha &=& \sqrt{\frac{p^4-1}{p^2 \log p}},
\nonumber\\
\beta &=& \sqrt{\frac{p^4-1}{2p^2 \log p}},
\een
we find that the matrices $K_{{\vec{n}}{\vec{m}}}^{uu}$ and $K_{{\vec{n}}{\vec{m}}}^{vv}$
become diagonal in the mode space
\ben
K_{{\vec{n}}{\vec{m}}}^{uu} &=& \frac{p^2}{p+1} \left( \prod_{i=1}^d  p^{\frac{p^2+1}{p^2-1}} \sqrt{\pi
\frac{p^2 \log p}{p^4-1}}\, 2^{n_i} {n_i}! \right) \left( p^{2\sum n_i-\frac{1}{4}\Box_y}
-p^2 \right) \delta_{{\vec{n}}{\vec{m}}},
\nonumber\\
K_{{\vec{n}}{\vec{m}}}^{vv} &=& \lambda^2 \left( \prod_{i=1}^d  p^{\frac{p^2+p+1}{p^2-1}} \sqrt{\pi \frac{2
p^2 \log p}{p^4-1}}\, 2^{n_i} {n_i}! \right) \left( p^{2\sum n_i-\frac{1}{2}\Box_y} -p
\right) \delta_{{\vec{n}}{\vec{m}}}.
\een
With the help of the approximate factorization (\ref{detfactor}) we find that
the spectrum of fluctuations naturally splits into two series
\ben
m_{\delta\psi}^2 &=& 8(N-1) + O(\lambda^2),
\nonumber\\
m_{\delta\phi}^2 &=& 4N-2 + O(\lambda^2),
\een
where $N=n_1+\cdots+n_d$ is the total oscillation number. It strongly resembles
the spectrum of string theory. At zero level $(N=0)$ we have two tachyons with
$m^2$ equal to $-2$ or $-8$. These are the open and closed string tachyon
respectively. Then there are $d$ massless modes, which for small but finite
$\lambda$ are just the Goldstone modes translating both lumps simultaneously in
the transverse directions. There is obviously no Goldstone mode which could be
associated with the open string spectrum and which would make the D-brane leave
the spacetime it sits in. The higher level states in the spectrum have the same
spacing as for the open or closed string.

All the fluctuation modes can be viewed as fields propagating on our closed
tachyon lump. These lumps look in all respects as spacetimes of lower
dimensions, whose dynamics is governed by some noncritical string theory. The
fact that we cannot put on these closed lumps open lumps of higher dimension,
and that the excitation modes of the open lumps we can put in cannot escape in
the transverse direction, is only confirming our interpretation.

The  statement that the fluctuations of the background are confined to the lump
is tantamount to the fact that $\delta\phi(y,x)$ are normalizable in the $x$
direction. Further support to our claim that none of the modes of the p-adic
closed string propagates outside the closed tachyon  lump, comes from numerical
study\footnote{ To solve (\ref{eom}) numerically, we use iterations of the
convolution formulae (see \cite{Moeller-Zwiebach} and \cite{Volovich}),
starting with the solution at zero coupling (\ref{double1}) and
(\ref{double2}).
 For lumps, however,
the iteration method by itself is not convergent. But we can fix this
by rescaling the solutions at each step: $\phi(x) \rightarrow \alpha
\phi(x)$ and $\psi(x) \rightarrow \beta \psi(x)$, where $\alpha$ and
$\beta$ can be determined from the integral over $x$ of the equations
of motion (\ref{eom}).}
of the double lump solution at finite coupling $\lambda$.
\begin{figure}[!ht]
\begin{center}
\input{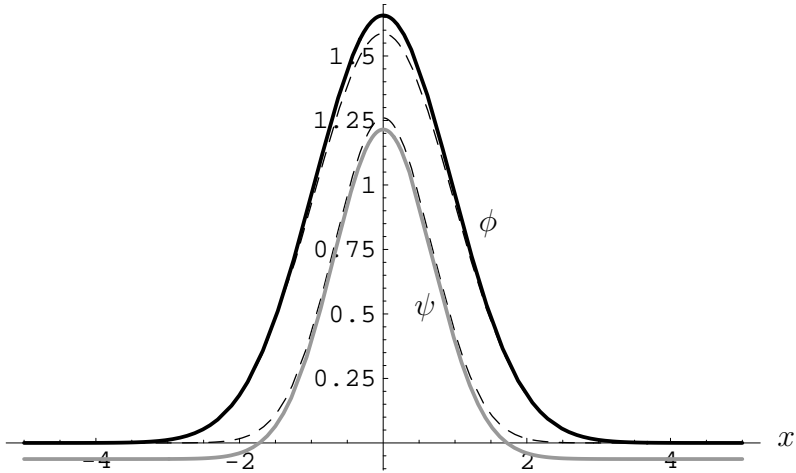}
\caption{Numerical solution for the double lump of codimension one,
with $p=2$ and $\lambda = 1/2$. The thick black line is $\phi(x)$ and
the thick gray line is $\psi(x)$. For comparison, we show, in dashed
lines, the solutions for $\lambda = 0$. We note that $\psi(x)$ tends
to its asymptotic value very fast, This is in contrast to the case of
the single lump of $\phi$, which causes oscillations of $\psi$ (see
Section \ref{s_back_reaction}). This is related to the fact that
closed string modes do not propagate outside the double lump.}
\label{f_double_lump}
\end{center}
\end{figure}
It is clear on our numerical solution in Fig. \ref{f_double_lump} for $p=2$ and
$\lambda = 1/2$ that both fields approach their true vacuum very fast.
This is unlike the case of back-reaction of a static D-brane in the perturbative
closed string vacuum $\psi = 1$, which causes spatial oscillations of the closed string tachyon
far away from the lump as we will show in Section~\ref{s_back_reaction}.

\sectiono{Back-reaction of static D-branes}
\label{s_back_reaction}
In this section we shall study how static open string lumps backreact on the closed string
tachyonic background. Later in section \ref{BackreactionOfTheRolling} we will generalize
these results to decaying time dependent lumps.
In general finding an exact solution to the full equations of motion of
a realistic open-closed string field theory is a rather hard problem. In p-adic string theory
the full equations of motion (\ref{eom})
are much simpler, but still too complicated to admit a simple analytic solution for
finite values of $\lambda$, which is proportional to the string
coupling $g$. We shall start by writing a perturbative expansion for the open and closed
tachyon fields
\ben
\psi &=& \psi_0 + \lambda^2 \psi_1 + O(\lambda^4),
\\
\phi &=& \phi_0 + \lambda^2 \phi_1 + O(\lambda^4).
\een
As our starting point we take a solution describing a codimension-$d$ D-brane in a flat
background
\ben
\psi_0(x,y) &=& 1,
\\
\phi_0(x,y) &=& p^{\frac{d}{2(p-1)}} \, e^{-\frac{1}{2 \log p} \frac{p-1}{p} x^2}
\een
and we will look for the first order correction $\psi_1$ describing the back-reaction.
Here $x$ is the $d$-dimensional transverse coordinate to the D-brane and
$y$ is the coordinate along its worldvolume.
To find $\phi_0(x,y)$ we have made use of the identity (\ref{identity}).
The equation for $\psi_1$ looks as
\be\label{psi1eq}
p^{-\frac{1}{4} \Box} \psi_1= p^2 \psi_1 +
\frac{p-1}{2 p}\left(\phi_0^{p+1}-1\right).
\ee
Solutions to the homogeneous part of this equation are plane waves $e^{i k x}$ with
\be
k^2 = 8 + \frac{4}{\log p} \cdot 2\pi i n
\ee
parametrized by an arbitrary integer $n$. These values of $k^2$ appear of course also
as poles in the Green's function, they give us the particle content of the theory. The lowest
mode has mass squared $m^2 =-8$ and is the closed string tachyon.
The higher modes have complex mass squared and are a bit mysterious. One can argue that
in the adelic theory \cite{Freund-Olson-Witten} these particles disappear and the true
closed string spectrum appears.
The only fact we will need, which is true for either theory
is that the only particle with mass $m$ lying on (or very close to) the imaginary axis is
the tachyon.

Since the unperturbed lumps are spherically symmetric in the transverse dimension, we concentrate
now only on spherically symmetric solutions of the homogeneous part of the equation. The solutions
can be readily found in terms of the Bessel functions, one of them being
\be
\psi_{1 \, homog.} \sim r^{-\frac{d-2}{2}} J_{\frac{d-2}{2}}\left(\sqrt{8} r\right).
\ee
A second solution can be obtained by changing the Bessel function $J_\nu$ into $Y_\nu$,
it is singular at the origin for $d>1$ however.
To find a particular solution of (\ref{psi1eq})
we first get rid of the constant part by writing
\be
\psi_{1}(x) = \frac{1}{2 p(p+1)} + \tilde \psi_{1}(x)
\ee
and pass to the Fourier transform
\be
\tilde \psi_{1}(x) = \int {d^d}k \,  e^{i k x} \tilde \psi_{1}(k).
\ee
Now the equation (\ref{psi1eq}) takes the form
\be
p^{\frac{1}{4} k^2}\tilde \psi_{1}(k) = p^2\tilde \psi_{1}(k)  +
\frac{p-1}{2 p} \left(\frac{p^{\frac{2p}{p-1}} \log p }{2\pi (p^2 -1)} \right)^{\frac{d}{2}}
e^{-\frac{p \log p}{2(p^2-1)} k^2}
\ee
and therefore a particular solution is given by
\be
\tilde \psi_{1}(x) =
\frac{p-1}{2 p} \left(\frac{p^{\frac{2p}{p-1}} \log p }{2\pi (p^2 -1)} \right)^{\frac{d}{2}}
\int {d^d}k \,  e^{i k x} \frac{1}{p^{\frac{1}{4} k^2}-p^2}\, p^{-\frac{p}{2(p^2-1)} k^2}.
\ee
While this integral is divergent on a codimension one hypersurface $k^2=8$,
it can be assigned a principal value.
Passing to the spherical coordinates, the integral can be rewritten as
\be\label{kint}
S_{d-2} \int_0^\infty dk\, k^{d-1} \int_0^\pi d\theta \,(\sin\theta)^{d-2} \,  e^{i k r \cos \theta}
\frac{1}{p^{\frac{1}{4} k^2}-p^2}\, p^{-\frac{p}{2(p^2-1)} k^2},
\ee
where $S_{d-2}$ is the volume of a $d-2$ dimensional sphere.
The integral over the angular variable is
\be
\int_0^\pi d\theta (\sin\theta)^{d-2} \,  e^{i k r \cos \theta} =
\sqrt{\pi}\, \Gamma\left(\frac{d-1}{2}\right) \left(\frac{2}{k r}\right)^{\frac{d-2}{2}}
J_{\frac{d-2}{2}}(k r),
\ee
the expression (\ref{kint}) now becomes
\be\label{kint2}
S_{d-2} \sqrt{\pi}\, \Gamma\left(\frac{d-1}{2}\right) \left(\frac{2}{r}\right)^{\frac{d-2}{2}}
\int_0^\infty dk\, k^{\frac{d}{2}}
J_{\frac{d-2}{2}}(k r)
\frac{1}{p^{\frac{1}{4} k^2}-p^2}\, p^{-\frac{p}{2(p^2-1)} k^2}.
\ee
To study the large $r$ behavior, we use the relation between the Bessel functions
$J_\nu(z)$, $Y_\nu(z)$ and the Hankel functions $ H_\nu^{(1,2)}$
\ben
J_\nu (z) &=& \frac{1}{2} \left( H_\nu^{(1)} +  H_\nu^{(2)}\right),
\\
Y_\nu (z) &=& \frac{1}{2i} \left( H_\nu^{(1)} -  H_\nu^{(2)}\right).
\een
Splitting the integral (\ref{kint2}) into two parts containing the Hankel functions
$ H_\nu^{(1)}$ and  $ H_\nu^{(2)}$, we add a small semi-circle contour to each of the principal
value integrals, such that in both cases the integration will be along a
single contour,
bypassing the pole at $k=\sqrt{8}$ from above or below respectively. These contours can then be further
deformed to straight lines going from the origin to infinity in a small finite angle
to the real axis, requiring only that the contours do not cross any of the other complex poles.
The integrals along these two lines are finite and regular and one can take a limit $r \to \infty$
in both of them separately. Using the known asymptotic behavior of the Hankel functions,
both integrals exhibit an exponential decay for large $r$. Therefore the only
exponentially unsuppressed contribution comes from the two small semi-circles around the pole.
Evaluating these half-residua we find
\be
S_{d-2} \sqrt{\pi}\, \Gamma\left(\frac{d-1}{2}\right) \left(\frac{2}{r}\right)^{\frac{d-2}{2}}
\,
Y_{\frac{d-2}{2}}(\sqrt{8} r) \,
\frac{(-\pi) 8^{\frac{d}{4}} }{\sqrt{2} p^2 \log p}\, p^{-\frac{4 p}{(p^2-1)}}.
\ee
Finally the most general spherically symmetric solution to (\ref{psi1eq}) behaves for
large $r$ as
\be\label{psi1}
\psi_{1}(r) \approx \frac{1}{2 p(p+1)} +
A r^{-\frac{d-2}{2}}
Y_{\frac{d-2}{2}}(\sqrt{8} r)  + B r^{-\frac{d-2}{2}}
J_{\frac{d-2}{2}}(\sqrt{8} r),
\ee
where
\be
A=-\frac{\pi}{\sqrt{8}}
\left(\frac{\sqrt{8}\,p^{\frac{2p}{p-1}} \log p }{ (p^2 -1)} \right)^{\frac{d}{2}}
\frac{p-1}{p^3 \log p}\, p^{-\frac{4 p}{(p^2-1)}}
\ee
and $B$ is an arbitrary constant. Note that the constant part in (\ref{psi1}) accounts
for the fact, that in the true open string vacuum with $\phi=0$, the value of the closed string
tachyon $\psi$ in its perturbative vacuum is shifted precisely by this constant times
$\lambda^2$ plus higher order corrections.

Now one may ask whether there are some natural boundary conditions on
the field $\psi(x)$, which would fix the value of the parameter $B$
and produce thus a unique answer for the perturbed background. For
large $r$ both the particular and the homogeneous solutions exhibit
the same kind of oscillations, they differ only in the relative phase
and by an arbitrary constant scale factor for the amplitude. There is
no way how one could fix uniquely this parameter. The physical
reason is that this ambiguity reflects the genuine instability of the
tachyonic background.

\sectiono{Time dependent solutions}
\label{s_time}

\label{BackreactionOfTheRolling}

We shall now turn our attention to the time dependent solutions of the coupled equations
of motion (\ref{eom}). Physically this is equivalent to the study
of closed strings emission from decaying D-branes
\cite{Okuda:2002yd,Chen:2002fp,Lambert:2003zr}.
We will be interested here in those solutions in which both open and
closed tachyon fields were in their perturbative vacua in the infinite past.
We will refer to those solutions as half-S-branes \cite{Strominger:2002pc}.
As in  \cite{Moeller-Zwiebach}  we are led to the following ansatz:
\ben\label{doublerollingansatz}
\phi(t) &=& 1- \sum_{n=1}^\infty a_n e^{\alpha n t},
\nonumber\\
\psi(t) &=& 1- \sum_{n=1}^\infty b_n e^{\beta n t},
\een
assuming that $\alpha,\beta >0$. The expansion in exponentials is very natural from the point
of view of string field theory \cite{Sen:2002nu}. One might expect at first, that
$\alpha=\sqrt{2}$ and $a_1 \ne 0$ describes open string tachyon condensation, whereas
$\beta=\sqrt{8}$ and $b_1 \ne 0$ describes closed string tachyon condensation.
Closer inspection reveals that indeed this is true to the zeroth order in the coupling $\lambda$.
At higher orders one has to take $\alpha=\beta$ and further equal to one of the two possible values
\ben\label{alpha12}
\alpha_1 &=& \sqrt{8} + \frac{p^2-1}{4 \sqrt{2}\, p^3 (1+p+p^2) \log p} \lambda^2 + O(\lambda^4),
\nonumber\\
\alpha_2 &=& \sqrt{2} - \frac{(p^2-1)(1+p\sqrt{p})}{4 \sqrt{2}\, p^{\frac{3}{2}}(1+p+p^2)  \log p}
\lambda^2 + O(\lambda^4).
\een
What might be more surprising is, that $a_1$ and $b_1$ cannot be chosen arbitrarily,
their ratio $x=b_1/a_1$ has to be fixed to one of the values
\ben
x_1 &=& 2(1+p+p^2) +  \frac{p^2-1}{p^3-1} \lambda^2 + O(\lambda^4),
 \nonumber\\
x_2 &=&  - \frac{p^2-1}{2 (p^3 - p^{\frac{3}{2}})} \lambda^2 + O(\lambda^4)
\een
respectively, depending on our choice of $\alpha$.
The authors of \cite{Moeller-Zwiebach} found a one parameter
family of solutions which corresponds to translations in time.
It has one parameter less than the Sen's solutions \cite{Sen:2002nu}, due to the fact that
we require our tachyon to be in its unstable vacuum in the far past.

For the two tachyon fields $\phi(t)$ and $\psi(t)$ one would expect
a two parameter family of solutions, but the solutions described by the ansatz
(\ref{doublerollingansatz}) have only one free parameter. Although it may look like we have
two discrete choices for $b_1/a_1$ corresponding to either open or closed tachyon condensation,
numerical investigation reveals that the two processes are mixed together.
\begin{figure}[t]
\begin{center}
\input{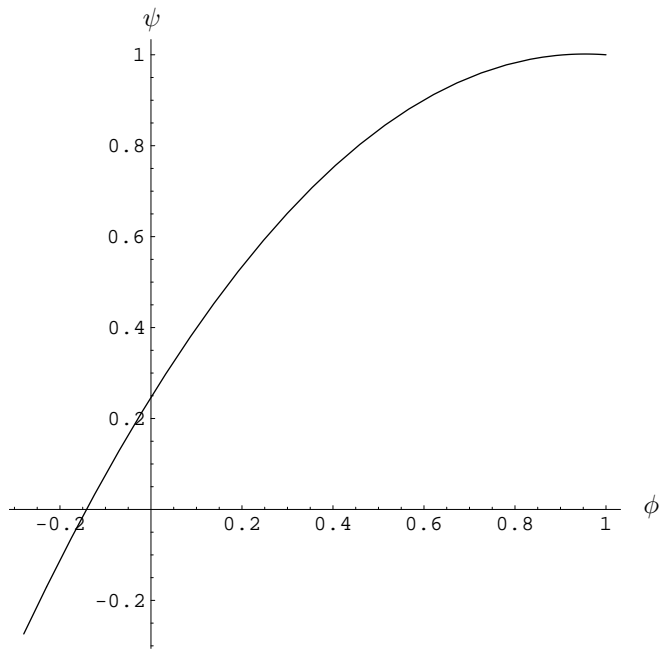}
\end{center}
\caption{Time evolution of the coupled open-closed tachyon system drawn in the $(\phi,\psi)$ plane.
The system starts its evolution in the infinite past at the point $(1,1)$, reaching the point
$\phi=0$ around $t \sim 0$. For this picture we have chosen $p=2$ and $\lambda = 0.001$,
to indicate that even at very  small coupling the exponential ansatz (\ref{doublerollingansatz})
leads to condensation of both tachyons.
}
\label{br_open_rolling}
\end{figure}
Solving recursively for the coefficients $a_n, b_n$ with the initial conditions corresponding
to the rolling of the open string tachyon, we find the time evolution of both fields $\phi(t)$
and $\psi(t)$, which we plot in the $\phi, \psi$ plane in Fig. \ref{br_open_rolling}.
Although at very early times the open string tachyon changes much faster than the closed one,
by the time the open string tachyon reaches zero, the closed one is also quite close to zero,
suggesting that actually both tachyons condense.
Following the evolution further on, beyond what is shown in Fig. \ref{br_open_rolling}, we find the wild
oscillations familiar from \cite{Moeller-Zwiebach}, whose significance is not yet entirely understood.

Another strategy for finding exact solutions to the coupled equations of motion would be using perturbation
theory as in Section \ref{s_back_reaction}. This will eventually lead to the correct two parameter
family of solutions. Let us describe first the solution where only the open string tachyon condenses.
We take $\psi_0=1$ and
\be
\phi_0(t)= 1 - e^{\sqrt{2} t} + \frac{1}{2(p^2+p+1)} e^{2\sqrt{2} t} + \cdots
\ee
being the rolling solution in open p-adic string theory studied in \cite{Moeller-Zwiebach}.
Solving for $\psi_1$ from (\ref{psi1eq}) we find
\be\label{psi1rollsol}
\psi_1(t) =\frac{p^2-1}{2p(p^2-\sqrt{p})} e^{\sqrt{2} t} +
\frac{(p^2-1)^2(p^2+1)}{4\sqrt{2}\,p^3(p^3-1)\log p}t  e^{2\sqrt{2} t}
+ \sum_{n=3}^\infty c_n e^{n \sqrt{2} t},
\ee
where $c_n$ are easily calculable $p$-dependent constants. We can of course add to (\ref{psi1rollsol}) an
arbitrary multiple of $e^{2\sqrt{2} t}$. What is a bit surprising, is the explicit appearance of
the factor $t$ in the second term of (\ref{psi1rollsol}). It stems from the fact that the closed string
tachyon mass is exactly twice the mass of the open one. For special values of the initial conditions, all
the terms at higher orders in $\lambda$ with explicit polynomial factors in $t$ combine into the
corrected exponentials of (\ref{alpha12}).
Note that those terms violate the periodicity in the imaginary time direction
\be
t \to t + \sqrt{2} \pi i.
\ee

One might think that the wild oscillations of both open and closed tachyon fields are related to the fact
that we were dealing with translationally invariant solutions, corresponding to the space-filling D-brane.
It is therefore interesting to look what happens
to the decay of a D0-brane, where one can imagine closed string ``radiation'' carrying away all the energy.
One can again use the ansatz (\ref{doublerollingansatz}) with $a_n$ and $b_n$ now being functions of $x$.
The equations for those coefficients are no longer algebraic, instead they contain factors like
$p^{-\Box/2}$. Analytically we can find the large distance behavior of the first order correction
in $\lambda$. This is analogous to calculating the emission of gravitational waves from a non static source
in general relativity.

Practically we are going to solve the equation (\ref{psi1eq}) with $\phi_0(x,t)$ being now a product
of the spatial lump $\phi_0^{lump}(x)$ and the rolling solution $\phi_0^{roll}(t)$
of \cite{Moeller-Zwiebach}. Expanding
$\psi_1(x,t)$ into powers of $e^{\sqrt{2}\, t}$ and solving for the leading large distance
behavior of the individual coefficients as in section \ref{s_back_reaction} we find
\be\label{psi1_inhom_rol}
\psi_{1}(r,t) \approx \frac{1}{2 p(p+1)} +
\sum_{n=0}^\infty A_n  e^{\sqrt{2}\, n t} r^{-\frac{d-2}{2}}
Y_{\frac{d-2}{2}}(r\, \sqrt{8-2n^2})  + B_n e^{\sqrt{2}\, n t} r^{-\frac{d-2}{2}}
J_{\frac{d-2}{2}}(r\, \sqrt{8-2n^2}),
\ee
where $A_n$ are finite calculable constants dependent on $p$ and $d$,
and $B_n$ are again free parameters.  We are getting thus true spatial
oscillations (up to a power in $r$) for $n=0,1$. Terms with $n>2$ are
superluminal radial waves with speeds ranging between $3/\sqrt{5}
\approx 1.34$ and $1$ the speed of light.  This should not be too
surprising since we are dealing with closed string tachyon waves. One
might hope, that resumming the series (\ref{psi1_inhom_rol}) one would
get rid of the wild oscillations in time, but that is unfortunately
beyond our capabilities to prove or disprove.

It would be interesting to have a well defined splitting of the energy
between the open and closed string sector. Whereas we can confirm that
the total energy is conserved for the solution shown on
Fig. \ref{br_open_rolling}, we are unable to split it in the two
sectors in any meaningful way.  This problem is certainly familiar
from general relativity, where one can achieve this goal under very
special circumstances. In that case one can argue, that the difference
between constant ADM energy and time dependent Bondi energy measures
the energy emitted in the gravitational waves.\footnote{For recent
application of these ideas to the study of closed string tachyon
condensation, see \cite{Gutperle:2002ki}.}  Both ADM and Bondi energy
are defined by integrals over the metric at spatial infinity or ${\cal
I}^+$ respectively. Since in p-adic string theory the closed string
tachyon has different properties than the metric in general relativity,
it seems hard to find any useful analogue of the Bondi energy in the
p-adic string theory.

\subsubsection*{Some other solutions}

In purely open p-adic string theory, it seems very unlikely that there
exist lump solutions in the time direction. In fact, it has been shown
in \cite{Moeller-Zwiebach} that, when $p$ is even, there can be no
monotonic lump (a tachyon rolling monotonically from a value $\phi=b$
towards the true vacuum until it reaches some value $a<b$, then
rolling back to $\phi=b$ monotonically).

Here, we want to show that in open-closed p-adic string theory, open
tachyon lumps in time do exist as long as we allow the closed tachyon
field to diverge in the far past and far future. Unfortunately we are
not able to solve the equations of motion with a nonzero
coupling\footnote{We can solve the equations of motion numerically
when the fields at infinity converge fast enough to constant values
(like in the case of the double lump). But here we will see that
$\psi(t)$ diverges at infinity, and we are not able to construct a
numerical solution.}, we will thus only consider $\lambda=0$.

The equations of motion (\ref{eom}) admit an interesting solution
obtained by Wick-rotating the spatial lump (\ref{double1})
\be \label{tlump_phi}
\psi(t) = p^{1 \over p^2-1} \, \exp \left({p^2-1
\over p^2 \log p} \, t^2 \right) \,.
\ee
Plugging this into (\ref{eom}) and trying the ansatz $\phi(t) =
A e^{-b^2 t^2}$, we find
\be \label{tlump_psi}
\phi(t) = p^{-{2p+1 \over 2
(p^2-1)}} \, \exp\left({-{p-1 \over 2 \log p} \, t^2}\right) \,.
\ee

For this solution, the open tachyon is in its true vacuum in the far
past and in the far future, and in the vicinity of its perturbative
vacuum only for a short time around $t=0$. One might be tempted to interpret
this solution as an S-brane \cite{Gutperle-Strominger}. We should be very
cautious though, because the solution requires rather special
``cosmological'' circumstances and moreover lacks the periodicity in imaginary time.

\sectiono{Discussions and conclusions}
\label{s_conclusions}

Although the relation of p-adic string theory to more conventional
string theories remains rather unclear, the theory has proven to be a
good toy model for studies of tachyon condensation. One of the
surprising features of the closed p-adic strings is that they look
very similar to the open p-adic strings. In fact, both theories
contain only a tachyon with a peculiar nonlocal interaction.  In the
case of open strings there are known lump solutions, whose tensions
and excitation spectra give them the interpretation of D-branes.  We
found analogous solutions in the closed string sector as well. Based
on their lump-like nature and excitation spectra, we have argued that
they correspond to spacetimes of lower dimensions. Outside the core
of the lump, the tachyon field is in its true vacuum and one can
easily check that there are indeed no propagating perturbative
degrees of freedom. Further support to our claims comes from the study
of possible open string lumps, {\it i.e.} D-branes, which can be put
in this closed string background. Far from the core of the closed
string lump, the open string tachyon has to be in its true vacuum.
Therefore the D-brane dimension has to be always smaller or equal to
that of the spacetime.

It is very tempting to speculate that similar closed string tachyon
lumps exist in realistic nonsupersymmetric string theories. All
physics including gravity would be localized on these lumps, and they
would thus form a viable alternative to compactification or large
extra dimensions scenarios.

The second main problem we dealt with is the dynamics of the open
string tachyon condensation.  The p-adic model is simple enough to
allow us to solve for the backreaction of open tachyon configurations
on the closed string background.  In particular, we have given the
analytic form of the fluctuations of the closed tachyon far away from
a static or decaying D-brane.  Although we hoped that a p-adic string
theory including the closed string sector would have nicer time
dependent solutions than pure open p-adic string theory, that doesn't
seem to be the case. Our rolling solutions, calculated in a series
expansion scheme, do in fact oscillate with diverging amplitudes. One
could still hope, however, that including higher orders in the
coupling constant might tame these oscillations. Further investigation
is needed on this issue.  A slightly disappointing fact which should
have been expected, is that we were not able to separate the
expression for the total energy into separate open and closed sectors;
such a separation would be useful to see the dissipation of the
rolling tachyon's energy into the closed modes.  This is certainly
reminiscent of the situation in general relativity.

\section*{Acknowledgments}
We are indebted to Barton Zwiebach for many useful discussions and early
collaboration on this paper. We thank also Ian Ellwood, Ami Hanany, Yoonbai
Kim, Alex Maloney and Joe Minahan for helpful conversations. This work was
supported in part by DOE contract \#DE-FC02-94ER40818.


\begin{thebibliography}{99}


\bibitem{Brekke}
L.~Brekke, P.~G.~Freund, M.~Olson and E.~Witten,
``Nonarchimedean String Dynamics,''
Nucl.\ Phys.\ B {\bf 302}, 365 (1988).


\bibitem{Brekke-Freund}
L.~Brekke and P.~G.~Freund,
``P-Adic Numbers In Physics,''
Phys.\ Rept.\  {\bf 233}, 1 (1993).


\bibitem{senconj}
A.~Sen,
``Descent relations among bosonic D-branes,''
Int.\ J.\ Mod.\ Phys.\  {\bf A14}, 4061 (1999)
[hep-th/9902105].
``Stable non-BPS bound states of BPS D-branes,''
JHEP {\bf 9808}, 010 (1998)
[hep-th/9805019].
``Tachyon condensation on the brane antibrane system,''
JHEP {\bf 9808}, 012 (1998)
[hep-th/9805170].
``SO(32) spinors of type I and other solitons on brane-antibrane
pair,''
JHEP {\bf 9809}, 023 (1998)
[hep-th/9808141].


\bibitem{Ghoshal:2000dd}
D.~Ghoshal and A.~Sen,
``Tachyon condensation and brane descent relations in p-adic string  theory,''
Nucl.\ Phys.\ B {\bf 584}, 300 (2000)
[arXiv:hep-th/0003278].


\bibitem{Joe_modes}
J.~A.~Minahan,
``Mode interactions of the tachyon condensate in p-adic string theory,''
JHEP {\bf 0103}, 028 (2001)
[arXiv:hep-th/0102071].


\bibitem{Freund-Olson-Witten}
P.~G.~Freund and M.~Olson,
``Nonarchimedean Strings,''
Phys.\ Lett.\ B {\bf 199}, 186 (1987).
P.~G.~Freund and E.~Witten,
``Adelic String Amplitudes,''
Phys.\ Lett.\ B {\bf 199}, 191 (1987).


\bibitem{Frampton-Nishino}
P.~H.~Frampton and H.~Nishino,
``Theory Of P-Adic Closed Strings,''
Phys.\ Rev.\ Lett.\  {\bf 62}, 1960 (1989).


\bibitem{Minahan-quantum}
J.~A.~Minahan,
``Quantum corrections in p-adic string theory,''
arXiv:hep-th/0105312.

\bibitem{Gutperle-Strominger}
M.~Gutperle and A.~Strominger,
``Spacelike branes,''
JHEP {\bf 0204}, 018 (2002)
[arXiv:hep-th/0202210].


\bibitem{Sen:2002nu}
A.~Sen,
``Rolling tachyon,''
JHEP {\bf 0204}, 048 (2002)
[arXiv:hep-th/0203211].

\bibitem{Sen:2002in}
A.~Sen,
``Tachyon matter,''
JHEP {\bf 0207}, 065 (2002)
[arXiv:hep-th/0203265].

\bibitem{Sen:2002vv}
A.~Sen,
``Time evolution in open string theory,''
JHEP {\bf 0210}, 003 (2002)
[arXiv:hep-th/0207105].


\bibitem{Moeller-Zwiebach}
N.~Moeller and B.~Zwiebach,
``Dynamics with infinitely many time derivatives and rolling tachyons,''
JHEP {\bf 0210}, 034 (2002)
[arXiv:hep-th/0207107].

\bibitem{Yang:2002nm}
H.~t.~Yang,
``Stress tensors in p-adic string theory and truncated OSFT,''
JHEP {\bf 0211}, 007 (2002)
[arXiv:hep-th/0209197].

\bibitem{Aref'eva:2003qu}
I.~Y.~Aref'eva, L.~V.~Joukovskaya and A.~S.~Koshelev,
``Time evolution in superstring field theory on non-BPS brane. I: Rolling  tachyon and energy-momentum conservation,''
arXiv:hep-th/0301137.

\bibitem{Fujita:2003ex}
M.~Fujita and H.~Hata,
``Time Dependent Solution in Cubic String Field Theory,''
arXiv:hep-th/0304163.





\bibitem{Okuda:2002yd}
T.~Okuda and S.~Sugimoto,
``Coupling of rolling tachyon to closed strings,''
Nucl.\ Phys.\ B {\bf 647}, 101 (2002)
[arXiv:hep-th/0208196].

\bibitem{Strominger:2002pc}
A.~Strominger,
``Open string creation by S-branes,''
arXiv:hep-th/0209090.

\bibitem{Chen:2002fp}
B.~Chen, M.~Li and F.~L.~Lin,
``Gravitational radiation of rolling tachyon,''
JHEP {\bf 0211}, 050 (2002)
[arXiv:hep-th/0209222].

\bibitem{Maloney:2003ck}
A.~Maloney, A.~Strominger and X.~Yin,
``S-brane thermodynamics,''
arXiv:hep-th/0302146.

\bibitem{Lambert:2003zr}
N.~Lambert, H.~Liu and J.~Maldacena,
``Closed strings from decaying D-branes,''
arXiv:hep-th/0303139.

\bibitem{Volovich}
Y.~Volovich,
``Numerical study of nonlinear equations with infinite number of  derivatives,''
arXiv:math-ph/0301028.

\bibitem{Gutperle:2002ki}
M.~Gutperle, M.~Headrick, S.~Minwalla and V.~Schomerus,
JHEP {\bf 0301}, 073 (2003)
[arXiv:hep-th/0211063].

\bibitem{Gaiotto-Itzhaki-Rastelli}
D.~Gaiotto, N.~Itzhaki, L.~Rastelli,
``Closed Strings as Imaginary D-branes,''
arXiv: hep-th/0304192.


\end{thebibliography}
\end{document}